\documentstyle[12pt,fleqn]{article}
\normalbaselines
\begin{document}
\bibliographystyle{unsrt}
\def\beee{\begin{equation}}
\def\eeee{\end{equation}}
\def\dggg{^{\dagger}}
\vbox{\vspace{6mm}}

\begin{center}
{\large \bf QUONS,\\
AN INTERPOLATION BETWEEN\\
BOSE AND FERMI OSCILLATORS}\\[7mm]
O.W. Greenberg\\
{\it Center for Theoretical Physics\\
Department of Physics and Astronomy\\
University of Maryland\\
College Park, MD~~20742-4111}\\[5mm]
Preprint number 93-097\\
\end{center}

\vspace{2mm}

\begin{abstract}
After a brief mention of Bose and Fermi oscillators and of particles
which obey other types of statistics, including intermediate statistics,
parastatistics, paronic statistics, anyon statistics and infinite statistics,
I discuss the statistics of ``quons'' (pronounced to rhyme with muons),
particles whose annihilation and creation
operators obey the $q$-deformed commutation relation (the quon algebra or
q-mutator) which interpolates between fermions and bosons. I emphasize that the
operator for interaction with an external source must be an effective Bose
operator in all cases.  To accomplish this for parabose, parafermi and quon
operators, I introduce parabose, parafermi and quon Grassmann numbers,
respectively.  I also discuss interactions of non-relativistic quons,
quantization of quon
fields with antiparticles, calculation of vacuum matrix elements of
relativistic
quon fields, demonstration of the TCP theorem, cluster decomposition, and
Wick's
theorem for relativistic quon fields, and the failure of local commutativity of
observables for relativistic quon fields.  I conclude with the bound on the
parameter $q$ for electrons due to the Ramberg-Snow experiment.
\end{abstract}

\section{Introduction}
I start by reviewing the (Bose) harmonic oscillator.  I want to emphasize that
the commutation relation,
\begin{equation}
[a_i,a_j^{\dagger}]_-\equiv a_ia_j^{\dagger}-a_j^{\dagger}a_i=\delta_{ij},
\label{bose}
\end{equation}
and the vacuum condition which characterizes the Fock representation
\begin{equation}
a_i|0 \rangle=0 \label{vac}
\end{equation}
suffice to calculate all vacuum matrix elements of polynomials in the
annihilation and creation operators.  The strategy is to move annihilation
operators to the right, picking up terms with a contraction of an annihilation
and a creation operator.  When the annihilation operator gets to the vacuum on
the right, it annihilates it.  For example,
\begin{eqnarray}
\langle 0|a_{i_1}a_{i_2} \cdots a_{i_n}a_{j_m}^{\dagger} \cdots
a_{j_2}^{\dagger} a_{j_1}^{\dagger} |0 \rangle = \delta_{i_nj_m}
\langle 0|a_{i_1}a_{i_2} \cdots a_{i_{n-1}}a_{j_{m-1}}^{\dagger} \cdots
a_{j_2}^{\dagger} a_{j_1}^{\dagger} |0 \rangle   \nonumber \\
+\langle 0|a_{i_1}a_{i_2} \cdots a_{i_{n-1}}a_{j_m}^{\dagger}a_{i_n}
a_{j_{m-1}}^{\dagger} \cdots
a_{j_2}^{\dagger} a_{j_1}^{\dagger} |0 \rangle.          \label{recur}
\end{eqnarray}
Continuing this reduction, it is clear that this vacuum matrix element
vanishes,
unless the set $\{i_1, i_2, \cdots, i_n \}$ is a permutation of the
set $\{j_1, j_2, \cdots, j_m \}$ (this includes $n=m$).  In particular, {\it no
relation is needed between
two $a$'s or between two $a^{\dagger}$'s}.  As we know, it turns out that
\begin{equation}
[a_i, a_j]_-=0=[a_i^{\dagger}, a_j^{\dagger}]_-,     \label{aa}
\end{equation}
but these relations are redundant in the Fock representation.  Also,
only the totally symmetric (one-dimensional) representations of the
symmetric (i.e., permutation) group  ${\cal S}_n$ occur.

To construct observables in the free theory we can use the number operator,
$n_k$, or the transition operator, $n_{kl}$,
\begin{equation}
n_k=n_{kk}=a_k^{\dagger}a_k,~~~n_{kl}=a_k^{\dagger}a_l.        \label{nnn}
\end{equation}
The commutation relation,
\begin{equation}
[n_{kl}, a_m^{\dagger}]_-=\delta_{lm}a_k^{\dagger},          \label{nna}
\end{equation}
follows from Eq.(\ref{bose}).  The number operator has
integer eigenvalues,
\begin{equation}
n_k (a_k^{\dagger})^{\cal N}|0 \rangle =
{\cal N}(a_k^{\dagger})^{\cal N}|0 \rangle.        \label{n1}
\end{equation}
Using $n_k$ and $n_{kl}$ we can construct the Hamiltonian,
\begin{equation}
H=\sum_k\epsilon_kn_k,              \label{h1}
\end{equation}
and other observables for the free theory.  The Hamiltonian obeys
\begin{equation}
[H, a_l^{\dagger}]_-=\epsilon_l a_l^{\dagger}.         \label{e1}
\end{equation}
Analogous formulas of higher degree
in the $a$'s and $a^{\dagger}$'s give interaction terms.

I want to pay special attention to couplings to external sources in the
quon theory; in preparation for that I write the external Hamiltonian
in the
Bose case,
\begin{equation}
H_{ext}=\sum_k (j_k^{\star} a_k+a_k^{\dagger}j_k),   \label{ext1}
\end{equation}
where $j_k$ is a c-number; i.e.,
\begin{equation}
[j_k,a^{\dagger}_l]_-=[j_k,j^{\star}_l]_-=0,~ {\rm etc}.  \label{c}
\end{equation}
This satisfies the commutation relation
\begin{equation}
[H_{ext}, a_l^{\dagger}]_-=j_l^{\star}.         \label{j1}
\end{equation}
Equations (\ref{e1}) and (\ref{j1}) state that $H$ and $H_{ext}$
are ``effective Bose operators'' in the context of a free theory with an
external source.  In particular, Eq. (\ref{e1}) and (\ref{j1}) imply
\begin{equation}
[H, a_{l_1}^{\dagger}a_{l_2}^{\dagger} \cdots a_{l_n}^{\dagger}]_-=
\sum_i \epsilon_i a_{l_1}^{\dagger}a_{l_2}^{\dagger} \cdots a_{l_n}^{\dagger}
\label{add1}
\end{equation}
and
\begin{equation}
[H_{ext}, a_{l_1}^{\dagger}a_{l_2}^{\dagger} \cdots a_{l_n}^{\dagger}]_-=
\sum_i
a_{l_1}^{\dagger}a_{l_2}^{\dagger}a_{l_{i-1}}^{\dagger}j_{l_{i}}^{\star}
a_{l_{i+1}}^{\dagger}\cdots a_{l_n}^{\dagger},
\label{add11}
\end{equation}
so the energy is additive for a system of free particles.
The general definition of an effective Bose operator is that the Hamiltonian
density commutes with the field when the points are separated by a large
spacelike distance,
\beee
[{\cal H}({\bf x}), \phi({\bf y})]_-\rightarrow 0,
|{\bf x}-{\bf y}| \rightarrow \infty.            \label{eff}
\eeee
This definition holds for all cases, including quons.

Everything I stated for the Bose harmonic oscillator can be repeated for the
Fermi oscillator, with obvious modifications. The commutation relation
Eq. (\ref{bose}) is replaced by the anticommutation relation;
\begin{equation}
[a_i,a_j^{\dagger}]_+\equiv a_ia_j^{\dagger}+a_j^{\dagger}a_i=\delta_{ij}
\label{fermi}
\end{equation}
that, together with the vacuum condition which characterizes the Fock
representation,
\begin{equation}
a_i|0 \rangle=0 \label{vacf},
\end{equation}
again suffices to calculate all vacuum matrix elements of multinomials in the
annihilation and creation operators.  For example,
\begin{eqnarray}
\langle 0|a_{i_1}a_{i_2} \cdots a_{i_n}a_{j_m}^{\dagger} \cdots
a_{j_2}^{\dagger} a_{j_1}^{\dagger} |0 \rangle = \delta_{i_nj_m} \langle 0|
\langle 0|a_{i_1}a_{i_2} \cdots a_{i_{n-1}}a_{j_{m-1}}^{\dagger} \cdots
a_{j_2}^{\dagger} a_{j_1}^{\dagger} |0 \rangle \nonumber \\
-\langle 0|a_{i_1}a_{i_2} \cdots a_{i_{n-1}}a_{j_m}^{\dagger}a_{i_n}
a_{j_{m-1}}^{\dagger} \cdots
a_{j_2}^{\dagger} a_{j_1}^{\dagger} |0 \rangle.          \label{recurf}
\end{eqnarray}
Continuing this reduction, it is clear that this vacuum matrix element
vanishes,
unless the set $\{i_1, i_2, \cdots, i_n \}$ is a permutation of the
set $\{j_1, j_2, \cdots, j_m \}$.  In particular, again {\it
no relation is needed between
two $a$'s or between two $a^{\dagger}$'s}.  As we know, it turns out that
\begin{equation}
[a_i, a_j]_+=0=[a_i^{\dagger}, a_j^{\dagger}]_+,     \label{aaf}
\end{equation}
but these relations again are redundant in the Fock representation.  Also, as
we
know, only the totally antisymmetric (one-dimensional) representations of the
symmetric group occur.

To construct observables in the free theory we again use the number operator,
$n_k$, or the transition operator, $n_{kl}$,
\begin{equation}
n_k=n_{kk}=a_k^{\dagger}a_k,~~~n_{kl}=a_k^{\dagger}a_l.        \label{nnnf}
\end{equation}
The commutation relation
\begin{equation}
[n_{kl}, a_m^{\dagger}]_-=\delta_{lm}a_k^{\dagger}          \label{nnaf}
\end{equation}
follows from the commutation relation Eq.(\ref{fermi}).  The number operator
again has integer eigenvalues; now, however, the number of particles in a
single
quantum state can only be zero or one, since Eq.(\ref{aaf}), which holds in the
Fock representation, implies $a_i^{\dagger 2}=0$,
\begin{equation}
n_k (a_l^{\dagger})^{\cal N}|0 \rangle =
\delta_{kl} {\cal N}(a_l^{\dagger})^{\cal N}|0 \rangle, {\cal N}=0,1.
\label{n1f}
\end{equation}
Using $n_k$ and $n_{kl}$ we again can construct the Hamiltonian,
\begin{equation}
H=\sum_k\epsilon_kn_k              \label{h1f}
\end{equation}
and other observables for the free theory.  The Hamiltonian obeys
\begin{equation}
[H, a_l^{\dagger}]_-=\epsilon_l a_l^{\dagger}.         \label{e1f}
\end{equation}
Analogous formulas of higher degree
in the $a$'s and $a^{\dagger}$'s give interaction terms.

I again pay special attention to couplings to external sources;
the external Hamiltonian in the Fermi case is
\begin{equation}
H_{ext}=\sum_k (f_k^{\star} a_k+a_k^{\dagger}f_k),   \label{ext1f}
\end{equation}
where $f_k$ is an anticommuting (Grassmann) number,
\begin{equation}
[f_k,f_l]_+=[f_k,f_l^{\star}]_+=[f_k,a_l]_+=[f_k,a_l^{\dagger}]_+=0. \label{g}
\end{equation}
The external Hamiltonian satisfies the commutation relation,
\begin{equation}
[H_{ext}, a_l^{\dagger}]_-=f_l^{\star}.         \label{j1f}
\end{equation}
The commutation relations Eq.(\ref{e1f}) and Eq.(\ref{j1f})
state that $H$ and $H_{ext}$
are ``effective Bose operators'' in the context of a free theory with an
external source.  In particular, Eq.(\ref{e1f}) and Eq.(\ref{j1f}) imply
\begin{equation}
[H, a_{l_1}^{\dagger}a_{l_2}^{\dagger} \cdots a_{l_n}^{\dagger}]_-=
\sum_i \epsilon_i a_{l_1}^{\dagger}a_{l_2}^{\dagger} \cdots a_{l_n}^{\dagger}
\label{add1f}
\end{equation}
and
\begin{equation}
[H_{ext}, a_{l_1}^{\dagger}a_{l_2}^{\dagger} \cdots a_{l_n}^{\dagger}]_-=
\sum_i
a_{l_1}^{\dagger}a_{l_2}^{\dagger}a_{l_{i-1}}^{\dagger}f_{l_{i}}^{\star}
a_{l_{i+1}}^{\dagger}\cdots a_{l_n}^{\dagger},
\label{add11f}
\end{equation}
so that the energy is additive for a system of free particles.

Notice that Eq.(\ref{vac},\ref{nnn},\ref{nna},\ref{h1},\ref{e1},\ref{add1}) for
the Bose case are {\it identical} to
Eq.(\ref{vacf},\ref{nnnf},\ref{nnaf},\ref{h1f},\ref{e1f},\ref{add1f})
for the Fermi case.
Eq.(\ref{n1},\ref{ext1},\ref{j1},\ref{add11}) for the
Bose case are analogous to Eq.(\ref{n1f},\ref{ext1f},\ref{j1f},\ref{add11f})
for the Fermi case.  Finally,
Eq.(\ref{bose},\ref{recur},\ref{aa},\ref{c}) for the Bose case and
Eq.(\ref{fermi},\ref{recurf},\ref{aaf},\ref{g})
for the Fermi case differ only by minus signs.

\section{Generalizations of Bose and Fermi Statistics}
As far as I know, the first attempt to go beyond Bose and Fermi statistics was
made by G. Gentile \cite{ge}.  He suggested ``intermediate statistics,''
in which up to $n$ particles can occupy  a given quantum state.  Clearly Fermi
statistics is recovered for $n=1$ and Bose statistics is recovered in the limit
$n \rightarrow \infty$.  As formulated by Gentile, intermediate statistics is
not a proper quantum statistics, because the condition of having at most $n$
particles in a quantum state is not invariant under change of basis.

H.S. Green \cite{hs} invented a generalization which is invariant under change
of basis.  I later dubbed his invention ``parastatistics'' \cite{dgs}.  Green
noticed that the number operator and transition operator,
Eq.(\ref{nnn}, \ref{nnnf}), have the same form
for both bosons and fermions, as do the commutation relations between the
transition operator and the creation and annihilation operators, Eq.(\ref{nna},
\ref{nnaf}).  Green generalized the transition operator
by writing
\begin{equation}
n_{kl}=(1/2)([a^{\dagger}_k,a_l]_{\pm} \mp p \delta_{kl}), \label{nbf}
\end{equation}
where the upper signs are for the generalization of bosons (``parabosons'') and
the lower signs are for the generalization of fermions (``parafermions'').
Since Eq.(\ref{nbf}) is trilinear, two conditions the states are necessary to
fix the Focklike representation:  the usual vacuum condition is
\begin{equation}
a_k|0\rangle=0;         \label{vacp}
\end{equation}
the new condition
\begin{equation}
a_ka^{\dagger}_l|0\rangle=p~ \delta_{kl}, p~~ {\rm integer},   \label{1p}
\end{equation}
contains the parameter $p$ which is the order of the parastatistics.
The Hamiltonian for free particles obeying parastatistics has the same form,
in terms of the number operators, as for Bose and Fermi statistics,
\begin{equation}
H=\sum_k \epsilon_k n_k,~ {\rm where,~ as~ usual}~
[H,a^{\dagger}_l]_-=\epsilon_l a^{\dagger}_l.   \label{pcr}
\eeee
For interactions with an external source, we must introduce para-Grassmann
numbers which make the interaction Hamiltonian an effective Bose operator.
This
is in analogy with the cases of external Bose and Fermi sources discussed
above.
We want
\beee
[H_{ext}, a^{\dagger}_l]=c^{\star}_l .   \label{pext}
\eeee
We accomplish this by choosing $H_{ext}=\sum_{kl}H^{ext}_{kl}$, with
\beee
H^{ext}_{kl}=(1/2)([c^{\star}_k, a_l]_{\pm}+
[a\dggg_k,c_l]_{\pm}),  \label{hpara}
\eeee
where the para-Grassmann numbers $c_k$ and $c\dggg_k$ obey
\beee
[[c^{\star}_k,c_l]_{\pm},c^{\star}_m]_-=0,~
[[c^{\star}_k,a_l]_{\pm},a\dggg_m]_-
=2 \delta_{lm}c^{\star}_k,~{\rm etc.},  \label{pgrass}
\eeee
and the upper (lower) sign is for parabose-Grassmann (parafermi-Grassmann)
numbers.  The ``etc.'' in Eq.(\ref{pgrass}) means that when some of the $c$'s
or $c\dggg$'s is replaced by an $a$ or an $a\dggg$, the relation retains its
form, except when the $a$ and $a\dggg$ can contract, in which case the
contraction appears on the right-hand-side.

It is worthwhile to make explicit the fact that in theories with
parastatistics states belong to many-dimensional representations of the
symmetric group.  This
contrasts with the cases of Bose and Fermi statistics in which only the one
dimensional representations occur.

I emphasize that parastatistics is a perfectly consistent local quantum field
theory.  The observables, such as the current, are local provided the proper
symmetrization or antisymmetrization is used; for example,
\beee
j^{\mu}(x)=(1/2)[{\bar \psi}(x),\gamma^{\mu}\psi(x)]_-   \label{1}
\eeee
for the current of a spin-1/2 field.  Further, all norms in a parastatistics
theory are positive; there are no negative probabilities.  On the other hand,
parastatistics of order $p>1$ gives a gross violation of statistics; for
example,
for a parafermi theory of order $p>1$ each quantum state can be occupied $p$
times.  A precise experiment is not needed to rule out such a gross violation.

Within the last three years two new approaches to particle statistics
(in three or more space dimensions) have
been studied in order to provide theories in which the Pauli exclusion
principle (i.e., Fermi statistics) and/or Bose statistics can be violated by a
small amount.  One of these approaches uses deformations
\cite{kn:ig,kn:gr,kn:gre,kn:go,kn:gree,kn:green,kn:ign,kn:ga,kn:bi,kn:igna}
of the trilinear
commutation relations of H.S. Green \cite{hs} and Volkov. \cite{vo}.
(Deformations of algebras and groups, in particular, quantum groups,
are a subject of great interest and activity at present.  The extensive
literature on this subject can be traced from \cite{kn:quantum}.)
The particles, called ``parons,'' which obey
this type of statistics have a quantum
field theory which is local,
but some states of such theories must have negative squared norms (i.e.,
there are negative probabilities in the theory).
The negative squared norms
first appear in many-particle states: in the model considered in
\cite{kn:gr} the first negative norm occurs in the state with Young tableau
(3,1).
It does not seem that the negative norm states decouple from those with
positive squared norms (as, in contrast, the corresponding states
do decouple in
manifestly covariant quantum electrodynamics).  Thus theories with parons seem
to have a fatal flaw.

The other approach uses
deformations of the bilinear Bose and Fermi commutation relations
\cite{ggg,owg,rnm,za,bo,fi,owgr}.
The particles which obey this type of statistics, called "quons," have
positive-definite squared norms for a range of the deformation parameter, but
the observables of such theories fail to have the desired locality properties.
This failure raises questions about the validity of relativistic quon theories,
but, in contrast to the paron theories,
does not cause a problem with non-relativistic quon theories.  (As I prove
below, the TCP theorem and clustering hold for free relativistic quon
theories, so even relativistic quon theories may be interesting.)

Still other approaches to violations of statistics were given in
\cite{kn:ok,kn:ra,kn:ram}.
An interpolation between Bose and Fermi
statistics using parastatistics of increasing order was studied in
\cite{kn:high}; this also does not give a small violation.

Yet another type of statistics, anyon statistics, has been extensively
discussed
recently, and applied to the fractional Hall effect and to high-temperature
superconductivity.  For anyons, the transposition of two particles can give
{\it any} phase,
\beee
\psi(1,2)=e^{i \phi} \psi(2,1).   \label{4}
\eeee
In the form usually considered, anyons only exist in two
space dimensions, and are outside the framework I am considering.  I will not
discuss them further here; rather I give two relevant references \cite{lm,wil}.

\section{The Quon Algebra}
\subsection{\it The $q=0$ case}
In their general classification of possible particle statistics, Doplicher,
Haag
and Roberts \cite{dhr} included bosons, fermions, parabosons, parafermions and
one other case, infinite statistics, in which all representations of the
symmetric group could occur, but did not give an algebra which led to this last
case.  Roger Hegstrom, a chemist at Wake Forest University, suggested averaging
the Bose and Fermi commutation relations to get
\beee
a_ka\dggg_l=\delta_{kl}, ~~a_k|0\rangle=0.   \label{5}
\eeee
(Unknown to Hegstrom and me, this algebra had been considered earlier by Cuntz
\cite{cun}.)
With Hegstrom's permission, I followed up his idea and showed that this algebra
gives an example of infinite statistics.  Consider a general scalar product,
\beee
(a\dggg_{k_1} \cdots a\dggg_{k_n}|0\rangle, a\dggg_{P^{-1}k_1} \cdots
a\dggg_{P^{-1}k_m}|0\rangle).                \label{6}
\eeee
This vanishes unless $n=m$ and $P$ is the identity, and then it equals one.
{}From this it follows that one can choose coefficients $c(P)$ to project into
states in each irreducible of $S_n$ and that the norm will be positive,
\beee
\|\sum_P c(P) a\dggg_{P^{-1}k_1} \cdots a\dggg_{P^{-1}k_n} |0\rangle \|^2 >0;
\eeee
thus every representation of $S_n$ occurs.  Note that there is
no relation between two $a$'s or two $a\dggg$'s; as before, the Fock vacuum
condition makes such relations unnecessary.

To construct observables, we want a number operator and a transition operator
which obey
\beee
[n_k,a\dggg_l]_-=\delta_{kl}a\dggg_l,~~[n_{kl},a\dggg_m]_-=\delta_{lm}a\dggg_k.
 \label{62}
\eeee
Once Eq.(\ref{62}) holds, the Hamiltonian and other observables can be
constructed in the usual way; for example,
\beee
H=\sum_k \epsilon_k n_k,~~ {\rm etc.}  \label{8}
\eeee
The obvious thing is to try
\beee
n_k=a\dggg_k a_k.   \label{9}
\eeee
Then
\beee
[n_k,a\dggg_l]_-=a\dggg_ka_ka\dggg_l-a\dggg_la\dggg_ka_k.  \label{10}
\eeee
The first term in Eq.(\ref{10}) is $\delta_{kl}a\dggg_k$ as desired; however
the second term is extra and must be canceled.  This can be done by adding the
term $\sum_ta\dggg_ta\dggg_ka_ka_t$ to the term in Eq.(\ref{9}).  This cancels
the extra term, but adds a new extra term, which must be canceled by another
term.  This procedure yields an infinite series for the number operator
and for the transition operator,
\beee
n_{kl}=a\dggg_ka_l+\sum_ta\dggg_ta\dggg_ka_la_t+\sum_{t_1,t_2}a\dggg_{t_2}
a\dggg_{t_1}a\dggg_ka_la_{t_1}a_{t_2}+ \dots   \label{11}
\eeee
As in the Bose case, this infinite series for the transition or number
operator defines an unbounded operator whose domain includes states made by
polynomials in the creation operators acting on the vacuum.
(As far as I know, this is the first case in which the number operator,
Hamiltonian, etc. for a free field are of infinite degree.)

For nonrelativistic theories, the $x$-space form of the transition operator is
\begin{eqnarray}
\rho_1({\bf x};{\bf y})=\psi\dggg({\bf x})\psi({\bf y})
+\int d^3z\psi\dggg
({\bf z})\psi\dggg({\bf x})\psi({\bf y})\psi({\bf z})  \nonumber \\
+\int d^3z_1d^3z_2\psi({\bf
z_2})\psi\dggg({\bf z_1})\psi\dggg({\bf x})\psi({\bf y})\psi({\bf z_1})
\psi({\bf z_2})+ \cdots,  \label{121}
\end{eqnarray}
which obeys the nonrelativistic locality requirement
\beee
[\rho_1({\bf x};{\bf y}),\psi\dggg({\bf w})]_-=\delta({\bf y}-{\bf
w})\psi\dggg({\bf x}),~~ {\rm and}~~
\rho({\bf x};{\bf y})|0\rangle=0.  \label{12}
\eeee
The apparent nonlocality of this formula associated with the space integrals
has
no physical significance.  To support this last statement, consider
\beee
[Qj_{\mu}(x),Qj_{\nu}(y)]_-=0,~~x \sim y,   \label{13}
\eeee
where $Q=\int d^ex j^0(x)$.  Equation (\ref{13}) seems to have nonlocality
because of the space integral in the $Q$ factors; however, if
\beee
[j_{\mu}(x),j_{\nu}(y)]_-=0,~~x \sim y, \label{14}
\eeee
then Eq.({\ref{13}) holds, despite the apparent nonlocality.  What is relevant
is the commutation relation, not the representation in terms of a space
integral.  (The apparent nonlocality of quantum electrodynamics in the Coulomb
gauge is another such example.)

In a similar way,
\beee
[\rho_2({\bf x}, {\bf y};{\bf y}^{\prime}, {\bf x}^{\prime}),
\psi\dggg({\bf z})]_-=\delta({\bf x}^{\prime}-{\bf z})
\psi\dggg({\bf x})\rho_1({\bf y},{\bf y}^{\prime})+
\delta({\bf y}^{\prime}-{\bf z})
\psi\dggg({\bf y})\rho_1({\bf x},{\bf x}^{\prime}).   \label{15}
\eeee
Then the Hamiltonian of a nonrelativistic theory with two-body interactions has
the form
\beee
H=(2m)^{-1} \int d^3x \nabla _x \cdot \nabla_{x^{\prime}}
\rho_1({\bf x}, {\bf x}^{\prime})|_{{\bf x}={\bf x}^{\prime}} +
\frac{1}{2} \int d^3x d^3y V(|{\bf x}-{\bf y}|)
\rho_2({\bf x},{\bf y};{bf y}, {\bf x}).              \label{16}
\eeee
\begin{eqnarray}
[H,\psi\dggg({\bf z}_1) \dots \psi\dggg({\bf z}_n)]_-=
-(2m)^{-1} \sum_{j=1}^n \nabla^2_{{\bf z}_i}+
 \sum _{i<j}
V(|{\bf z}_i-{\bf z}_j|)] \psi\dggg({\bf z}_1) \dots \psi\dggg({\bf z}_n)
\nonumber \\
+\sum_{j=1}^n \int d^3x V(|{\bf x}-{\bf z}_j|)\psi\dggg({\bf z}_1)
\cdots \psi\dggg({\bf z}_n)\rho_1({\bf x}, {\bf x}^{\prime}).   \label{17}
\end{eqnarray}
Since the second term on the right-hand-side of Eq.(\ref{17}) vanishes when the
equation is applied to the vacuum, this equation shows that the usual
Schr\"odinger equation holds for the $n$-particle system.  Thus the usual
quantum mechanics is valid, with the sole exception that any permutation
symmetry is allowed for the many-particle system.  This construction justifies
calculating the energy levels of (anomalous) atoms with electrons in states
which violate the exclusion principle using the normal Hamiltonian,
but allowing anomalous
permutation symmetry for the electrons {\cite{drake}.

In general, an arbitrary many-particle state is in a mixture of inequivalent
irreducible representations of $S_n$.  If ${\cal O}$ is any observable and
$\Psi$ is any state, the cross terms between irreducibles in the
matrix element $\langle \Psi |{\cal O}| \Psi \rangle$ automatically vanish,
since observables keep states inside their irreducible representation of $S_n$.
\subsection{\it The general quon algebra for $-1 \leq q \leq 1.$}
The quon algebra,
\beee
a_k a^{\dagger}_l-q a^{\dagger}_l a_k=\delta_{kl},         \label{p}
\eeee
which is a deformation of the Bose and Fermi algebras and interpolates between
these algebras as $q$ goes from $1$ to $-1$ on the real axis, shares many
qualitative features with the special case of $q=0$ just discussed.  In
particular, the quon algebra also allows all representations of $S_n$.
This algebra, supplemented by the vacuum condition
\beee
a_k|0\rangle=0,                                           \label{q}
\eeee
determines a (Fock-like) representation in a linear vector space.
For $-1<q<1,$
the squared norms of all vectors made by limits of polynomials of the creation
operators, $a^{\dagger}_k,$ are strictly positive\cite{za,bo,fi}.
Among other things, this
means that there are $n!$ linearly independent states of $n$ particles with
distinct quantum numbers, and all representations of the symmetric group occur.
Also, as in the case of $q=0$,
Eqs.(\ref{p},\ref{q}) allow the
calculation of the vacuum to vacuum matrix element
of any polynomial in the $a$'s and
$a^{\dagger}$'s.  As before, no commutation relation
between two $a$'s or between
two $a\dggg$'s is needed.  Further, in this case, no such rule can be imposed
on $aa$ or $a^{\dagger}a^{\dagger}$.
The relation,
\beee
a_ka_l-qa_la_k=0,                                  \label{444}
\eeee
between two
$a$'s which one might guess
in analogy with the Bose and Fermi commutation rules holds only when $q^2=1$;
and requires that $q=\pm 1$ in Eq.(\ref{p}); i.e., Eq.(\ref{444}) can hold only
in the Bose and Fermi cases.  To see this,
interchange $k$ and $l$ in Eq.(\ref{444}) and put the result
back in the initial relation.
(Commutation relations between two $a$'s or between two $a^{\dagger}$'s
are also not needed
for normal ordering, i.e., to expand a product of $a$'s and $a^{\dagger}$'s
as a sum of terms in which creation operators always stand to the left of
annihilation operators.  Wick's theorem for quon operators is similar to the
usual Wick's theorem; the only difference is that the terms acquire powers of
$q.$  I gave the precise algorithm in \cite{pr43}.)  As $q$ approaches $-1$
from above, the more antisymmetric representations become more heavily weighted
and at $-1$ only the antisymmetric representation survives.  As $q$ approaches
$1$ from below, the more symmetric representations become more heavily weighted
and at $1$ only the symmetric representation survives.  Outside the interval
$[-1,1]$, the squares of some norms become negative.

Now I discuss the construction of observables both without and with an
external source.  Without an external
source, one again needs a set of number operators $n_k$ such that
\beee
[n_k,a\dggg_l]_-=\delta_{kl}a\dggg_l.           \label{81}
\eeee
Like the $q=0$ case, the expression for $n_k$ or $n_{kl}$ is an infinite series
in creation and annihilation operators; unlike the $q=0$ case, the coefficients
are complicated.  The first two terms are
\beee
n_{kl}=a^{\dagger}_ka_l
+ (1-q^2)^{-1} \sum_t (a^{\dagger}_t a^{\dagger}_k
-q a^{\dagger}_ka^{\dagger}_t)(a_la_t-qa_ta_l) + \cdot \cdot \cdot. \label{221}
\eeee
Here I gave the transition number operator $n_{kl}$ for $k \rightarrow l$
since this takes no extra effort.  The general formula for the number operator
is given in \cite{stan} following a conjecture of Zagier \cite{za}.
As before, the Hamiltonian is
\beee
H=\sum_k \epsilon_k n_k,~~{\rm with}~~[H,a\dggg_l]_-=\epsilon_la\dggg_l.
                \label{82}
\eeee
For an external source, it is crucial to
insure that $H_{ext}$ is an effective Bose operator.  In order to do this, one
must choose the external source to be a quon analog of a Grassmann number,
i.e.,
to obey
\beee
c_k c^{\star}_l-qc^{\star}_lc_k=0;~c_k a\dggg_l-qa\dggg_l c_k=0;~
a_kc^{\star}_l-qc^{\star}_la_k=0.              \label{820}
\eeee
Then $H_{ext}$ must be chosen to obey
\beee
[H_{ext},a\dggg_l]_-=c^{\star}_l               \label{821}
\eeee
For example, for
$q=0$, the first two terms are
\beee
H_{ext}=\sum_k
(c^{\star}_ka_k+a\dggg_kc_k)+\sum_k\sum_ta\dggg_t(c^{\star}_ka_k+a\dggg_kc_k)a_t
+ \cdots                                   \label{85}
\eeee
For the general case, I give the first two terms of $H^{ext}_{kl}$, subject to
\beee
[H^{ext}_{kl}, a\dggg_m]_-=\delta_{lm}c^{\star}_k         \label{8001}
\eeee
and hermiticity, $(H^{ext}_{kl})\dggg=H^{ext}_{lk}$,
\begin{eqnarray}
H^{ext}_{kl}=c^{\star}_ka_l+a\dggg_kc_l+(1-q^2)^{-1}\sum_t(a\dggg_tc^{\star}_k-q
c^{\star}_ka\dggg_t)(a_la_t-qa_ta_l)     \nonumber          \\
+\sum_k(1-q^2)^{-1}(a\dggg_ta\dggg_k-qa\dggg_ka\dggg_t)(c_la_t-qa_tc_l) +
\cdots
           \label{8002}
\end{eqnarray}
If, instead, we incorrectly choose
$H_{ext}=\sum_k(j^{\star}_ka_k+a\dggg_kj_k)$,
where $j$ is a $c$-number, then the energy of widely separated states is not
additive,
\beee
H_{ext}a\dggg_{k_1}a\dggg_{k_2} \cdots a\dggg_{k_n}|0\rangle=
[j^{\star}_{k_1}a\dggg_{k_2} \cdots a\dggg_{k_n}+qa\dggg_{k_1}j^{\star}_{k_1}
\cdots a\dggg_{k_n} + \cdots q^{n-1} a\dggg_{k_1}a\dggg_{k_2} \cdots
j^{\star}_{k_n}]|0\rangle            \label{86}
\eeee
Although this point is transparent for the case of fermions where powers of
negative one replace powers of $q$ in Eq.(\ref{86}), it seems to be less clear
in
the quon case.  Because this point was not recognized, the bound on validity of
Bose statistics for photons given in \cite{fivel} is incorrect.

Again one- and two-body observables can be constructed from
$\rho_1({\bf x},{\bf x})$ and
$\rho_2({\bf x}_1,{\bf x}_2;{\bf y}_2,{\bf y}_1)$.  The formula for $n$ can be
translated into a formula for $\rho_1$, and at least the first non-trivial term
is known for $\rho_2$.  With these, a valid nonrelativistic theory of identical
particles with (small) violations of Fermi of Bose statistics can be formulated
\cite{owg2}.

The condition that observables must be effective
Bose operators leads to conservation of statistics which states that all
interactions must involve an even number of fermions or para-fermions
and an even number of para particles (except for cases in which $p$ para
fields can occur when the order of the parastatistics is
$p$)\cite{kn:greemess}.  I expect that
conservation of statistics must also hold for quons and, in particular,
that a single quon
cannot couple to normal fields \cite{kn:sud}.
I plan to discuss the conservation of
statistics for quons in detail elsewhere.
I have discussed the simple case of a single oscillator elsewhere\cite{pr43},
so I will not repeat this discussion here.

To summarize, all
irreducible representations of $S_n$ have positive (norm)$^2$ in this
interval.  As $q \rightarrow \pm 1$ the more symmetric (antisymmetric)
irreducibles occur with higher weight.  At the endpoints, $q= \pm 1,$ only
the symmetric (antisymmetric) representation survives.

\section{The quon algebra in the presence of antiparticles}
The pattern is established by discussing the spin-zero case.  Since
\beee
\phi(x)=\frac{1}{(2\pi)^{3/2}}\int\frac{d^3k}{\sqrt {2 \omega_k}}
(b_k e^{-ik \cdot x}+d^{\dagger}_ke^{ik \cdot x}),        \label{320}
\eeee
\beee
\phi^{\dagger}(x)=\frac{1}{(2\pi)^{3/2}}\int\frac{d^3k}{\sqrt {2 \omega_k}}
(d_k e^{-ik \cdot x}+b^{\dagger}_ke^{ik \cdot x}),         \label{330}
\eeee
$\omega_k=k^0=\sqrt{{\bf k}^2+m^2},$ to preserve charge conjugation symmetry
one
should supplement the commutation relation for the $b$'s and $b\dggg$'s by
\beee
d_kd\dggg_l-qd\dggg_ld_k=\delta_{kl},                    \label{340}
\eeee
\beee
d_kb\dggg_l-qb\dggg_ld_k=0,                             \label{350}
\eeee
With this choice, the positivity of the norms is preserved in the presence of
antiparticles.  If, instead, one chooses the $x$-space relation,
\beee
\phi(x)\phi\dggg(y)-q\phi\dggg(y)\phi(x)=F(x-y)\equiv {\rm vev}(lhs)
\label{306}
\eeee
then one finds the usual quon commutation relation for the $b$'s, but
\beee
d_kd\dggg_l-q^{-1}d\dggg_ld_k=\delta_{kl}             \label{360}
\eeee
for the $d$'s.  Since Eq.(\ref{360}) gives positive norms only {\it outside}
$|q|<1$, this choice is inconsistent.  In \cite{gov,cou} this last choice has
been argued to imply breaking of charge conjugation invariance.

It is amusing to note that the TCP theorem and clustering hold, at least for
free quon fields, despite the
failure of locality \cite{pr43}.

\section{Experiments}
In a conference devoted to issues related to harmonic oscillators, it is
worthwhile to
make some comments about the experimental relevance of the quon oscillator.
The quon oscillator provides a parametrization of possible small departures
from
Fermi or Bose statistics.
The simplest way to detect small violations of statistics is to find a state
which either Fermi or Bose statistics would not allow.  For Fermi
(Bose) statistics,
this would be a state in which identical particles are not totally
antisymmetric (symmetric).
The path-breaking high-precision experiment of Ramberg and Snow\cite{kn:ramb}
searches for transitions to a state in which the electrons of the copper atom
are not totally antisymmetric.  The failure to detect such transitions (above
background)
leads to the following upper bound on violation of the exclusion principle,
\beee
\rho_2={1 \over 2}(1-\beta^2)\rho_a+{1 \over 2} \beta^2 \rho_s,
{}~~~~ {1 \over 2} \beta^2 \leq 1.7 \times 10^{-26},
\label{41}
\eeee
$\rho_2$ is the two-electron density matrix, $\rho_{a(s)}$ is the antisymmetric
(symmetric) two-electron density matrix.  For two electrons in different states
$\rho_2$ can be expressed in terms of $q$ of the q-mutator as
\beee
\rho_2={ 1 \over 2}(1-q) \rho_a+{1 \over 2}(1+q) \rho_s,          \label{42}
\eeee
so the Ramberg Snow bound is
\beee
0 \leq (1+q)/2 \leq 1.7 \times 10^{-26}.                          \label{43}
\eeee
A high-precision experiment to detect or bound violations of the exclusion
principle for electrons in helium is being conducted by D. Kelleher,
et al.\cite{kn:ke}

I conclude this brief discussion of experimental bounds on small violations
of statistics by remarking that there are three types of such
experiments: (1) to
detect an accumulation of particles in anomalous states, (2) to detect
transitions to anomalous states and (3) to detect deviations from the usual
statistical properties of many-particle systems.  Here and in
\cite{kn:gree} type (2) experiments are discussed, because they allow detection
of single transitions to anomalous states.  Type (1) experiments require
detection of a small concentration of anomalous states in a macroscopic system;
for that reason they are generally less sensitive than type (2) experiments.
I have not analyzed type (3) experiments; however it seems likely that they
will fail to provide high-precision tests for the same reason that type (1)
experiments fail: it will be difficult to detect the modification of the
statistical properties of a macroscopic sample due to a small concentration
of anomalous states.

\section{Summary}
The quon oscillator serves as an interpolation between Fermi and Bose
statistics.  This interpolation preserves positivity of norms and the
non-relativistic form of locality, but fails to allow local observables in a
relativistic theory.  Nonetheless, the TCP theorem and clustering hold in
relativistic quon theories.
Terms in the Hamiltonian for both self-interacting
systems and systems interacting with an external source must be effective Bose
operators in order for the additivity of the energy for widely separated
subsystems to hold.  The quon theory provides a parametrization of possible
deviations from Bose or Fermi statistics.
\vglue.2in

\section{Acknowledgements}
This work was supported in part by the National Science Foundation
and by a Semester Research Award from the University of Maryland, College
Park.

e-mail addresses: greenberg@umdhep (bitnet)\\
umdhep::greenberg (decnet)\\
greenberg@umdep1.umd.edu (internet)

\begin{thebibliography}{99}
\bibitem{fi}
D.I. Fivel, Phys. Rev. Lett. {\bf 65}, 3361 (1990); erratum {\bf 69}, 2020
(1992).  I have not checked that the erratum repairs the error in the proof of
positivity in the article.
\end{thebibliography}
\end{document}